\documentclass[prl,aps,twocolumn,superscriptaddress,showpacs]{revtex4}
\usepackage{graphicx}

\begin{document}

\title{       Dimerization versus Orbital Moment Ordering 
              in the Mott insulator YVO$_3$ }

\author {     Peter Horsch }
\affiliation{ Max-Planck-Institut f\"ur Festk\"orperforschung,
              Heisenbergstrasse 1, D-70569 Stuttgart, Germany }
\author {     Giniyat Khaliullin }
\affiliation{ Max-Planck-Institut f\"ur Festk\"orperforschung,
              Heisenbergstrasse 1, D-70569 Stuttgart, Germany }
\affiliation{ E. K. Zavoisky Physical-Technical Institute of the
              Russian Academy of Sciences, 420029 Kazan, Russia } 
\author {     Andrzej M. Ole\'{s} }
\affiliation{ Max-Planck-Institut f\"ur Festk\"orperforschung,
              Heisenbergstrasse 1, D-70569 Stuttgart, Germany }
\affiliation{ Departments of Theoretical Physics, Jagellonian
              University, Reymonta 4, PL-30059 Krak\'ow, Poland } 

\date{April 11, 2003}

\begin{abstract}
We use exact diagonalization combined with mean-field theory to 
investigate the phase diagram of the spin-orbital model for cubic 
vanadates. The spin-orbit coupling competes with Hund's exchange and 
triggers a novel phase, with the ordering of $t_{2g}$ orbital 
magnetic moments stabilized by the tilting of VO$_6$ octahedra. 
It explains qualitatively spin canting and reduction of magnetization 
observed in YVO$_3$. At finite temperature an orbital Peierls 
instability in the $C$-type antiferromagnetic phase induces modulation 
of magnetic exchange constants even in absence of lattice distortions.
The calculated spin structure factor shows a magnon splitting
due to the orbital Peierls dimerization.
 
\end{abstract}


\maketitle

Many transition metal oxides are Mott-Hubbard insulators, in which local
Coulomb interaction $\propto U$ suppresses charge fluctuations and leads 
to strongly correlated $3d$ electrons at transition metal ions  
\cite{Ima98}. When degenerate $d$ orbitals are partly filled, the 
{\it orbital degrees of freedom\/} have to be considered on equal 
footing with electron spins and the magnetic properties of undoped 
compounds are described by spin-orbital superexchange (SE) 
models \cite{Kug82,Tok00}. Such SE interactions are typically 
strongly frustrated on a perovskite lattice, leading to enhanced quantum 
effects \cite{Fei97}.  In systems with $e_g$ orbital degeneracy 
(manganites, cuprates) this frustration is usually removed by a 
structural transition that occurs well above the magnetic ordering 
temperature and lifts the orbital degeneracy via the cooperative 
Jahn-Teller (JT) effect. 

A different situation arises when $t_{2g}$ orbitals are partly filled 
like in titanium and vanadium oxides. As the JT coupling is much weaker, 
the intrinsic frustration between spin and orbital degrees of freedom 
may show up in this case. Unusual magnetic properties of titanates have 
recently been discussed in terms of coupled spin-orbital SE dynamics 
\cite{Kha00}. In addition to SE, spin and orbital occupancies of 
$t_{2g}$ levels are coupled also via atomic spin-orbit interaction, 
$H_{so}\propto\Lambda (\vec S\cdot\vec l)$, which is particularly
relevant for vanadates with a triplet $^3T_2$ ground state of V$^{3+}$ 
ions. As a result  of {\it intersite} SE and {\it on-site} 
$\Lambda$ interactions spin and orbital orderings/fluctuations are 
strongly coupled, as observed in the canonical spin-orbital system 
V$_2$O$_3$ \cite{Bao97}, as well as in cubic LaVO$_3$ \cite{Miy00}. 

The magnetic properties of YVO$_3$ are particularly puzzling 
\cite{Ren00}, and indicate dimerization along the FM direction within 
the $C$-AF phase \cite{Ulr02}. In this Letter we argue that such an 
exotic $C$-AF phase follows from the realistic spin-orbital model for 
vanadates that emphasizes the competition between SE bond physics and 
intraatomic spin-orbit coupling $\propto\Lambda$. We investigate the 
phase diagram of this model and show that {\it orbital moments\/} are 
induced in the $C$-phase by finite $\Lambda$, and form at larger 
$\Lambda$ a novel orbital moment (OM) phase.

The superexchange in cubic vanadates originates from virtual charge 
excitations, $d^2_id^2_j\rightarrow d^3_id^1_j$, by the hopping $t$ 
which couples pairs of identical orbitals. When such processes are 
analyzed on individual bonds $\langle ij\rangle\parallel\gamma$ along 
each cubic axis $\gamma=a,b,c$, one finds the spin-orbital Hamiltonian 
with $S=1$ spins ($J=4t^2/U$) \cite{Kha01},
\begin{equation}
\label{som}
{\cal H}=J\sum_{\gamma}\sum_{\langle ij\rangle\parallel\gamma}\left[ 
     \frac{1}{2}({\vec S}_i\cdot {\vec S}_j+1)
     {\hat J}_{ij}^{(\gamma)} + {\hat K}_{ij}^{(\gamma)}\right]
     +H_{so},
\end{equation}
where the orbital operators ${\hat J}_{ij}^{(\gamma)}$ and
${\hat K}_{ij}^{(\gamma)}$ depend on the pseudospin $\tau=1/2$ operators 
${\vec\tau}_i=\{\tau_i^x,\tau_i^y,\tau_i^z\}$, given by two orbital 
flavors active along a given direction $\gamma$. For instance, $yz$ and 
$zx$ orbitals are active along $c$ axis, and we label them as $a$ and 
$b$, as they lie in the planes orthogonal to these axes. The general 
form of the superexchange (\ref{som}) was discussed before, and we have 
shown that strong quantum fluctuations in the orbital sector provide a 
mechanism for the $C$-AF phase \cite{Kha01}. When $c$ ($xy$) orbitals 
are occupied ($n_{ic}=1$), as suggested by the electronic structure 
\cite{Saw96} and by the lattice distortions in YVO$_3$ 
\cite{Ren00,Bla01}, the electron densities in $a$ and $b$ orbitals 
satisfy the local constraint $n_{ia}+n_{ib}=1$. The interactions along 
the $c$ axis simplify then to:
\begin{eqnarray}
\label{jc}
{\hat J}_{ij}^{(c)}\!&=&\!
(1+2R)\Big({\vec\tau}_i\cdot {\vec\tau}_j\!+\!\frac{1}{4}\Big)\! 
-\! r\Big({\vec\tau}_i\otimes{\vec\tau}_j\!+\!\frac{1}{4}\Big) \!-\! R,         
    \\ 
\label{kc}
{\hat K}_{ij}^{(c)}\!&=&\!
  R\Big({\vec\tau}_i\cdot {\vec\tau}_j\!+\!\frac{1}{4}\Big)  
+ r\Big({\vec\tau}_i\otimes{\vec\tau}_j\!+\!\frac{1}{4}\Big);           
\end{eqnarray}
they involve the fluctuations of $a$ and $b$ orbitals 
$\propto{\vec\tau}_i\cdot {\vec\tau}_j$, and 
${\vec\tau}_i\otimes{\vec\tau}_j=
\tau_i^x\tau_j^x -\tau_i^y\tau_j^y +\tau_i^z\tau_j^z$, 
while the interactions along the $\gamma=a(b)$ axis depend on the 
static correlations $\propto n_{ib}n_{jb}$ ($n_{ia}n_{ja}$) only; 
for instance:
 \begin{eqnarray}
 \label{ja}
 {\hat J}_{ij}^{(a)}\!&=&
 \frac{1}{2}\Big[(1-r)(1+n_{ib}n_{jb})-R(n_{ib}-n_{jb})^2\Big],  \\ 
 \label{ka}
 {\hat K}_{ij}^{(a)}\!&=&\frac{1}{2}(R+r)(1+n_{ib}n_{jb}).
 \end{eqnarray}
The Hund's exchange $\eta=J_H/U$ determines the multiplet structure of
$d^3$ excited states which enters via the coefficients: 
$R=\eta/(1-3\eta)$ and $r=\eta/(1+2\eta)$. The pseudospin operators in 
Eqs. (\ref{jc})--(\ref{kc}) may be represented by Schwinger bosons:
$\tau_i^x= (a_i^{\dagger}b_i^{}+b_i^{\dagger}a_i^{})/2$, 
$\tau_i^y=i(a_i^{\dagger}b_i^{}-b_i^{\dagger}a_i^{})/2$, 
$\tau_i^z=(n_{ia}^{}-n_{ib}^{})/2$. 

The individual VO$_6$ octahedra are tilted by angle $\phi_i=\pm\phi$, 
which alternate along the $c$ axis \cite{Bla01}. 
As the $xy$ orbital is inactive, two components of the 
orbital moment $\vec l_i$ are quenched, while the third one (2$\tau_i^y$), 
parallel to the local axis of a VO$_6$ octahedron, couples to the spin 
projection. Because of AF correlations of $\tau_i^y$ moments, spin-orbit 
coupling induces a staggered spin component. As the spin interactions 
are FM, weak spin-orbit coupling would give no energy gain, if the spins 
were oriented along the $c$ axis. Thus, finite $\Lambda$ breaks the 
SU(2) symmetry and favors easy magnetization axis within the $(a,b)$ 
plane. 
As quantization axis for $\vec l_i$ ($\vec S_i$) we use the octahedral axis
(and its projection on the  $(a,b)$ plane), respectively.
The spin-orbit coupling in Eq. (\ref{som}) is 
then given by: 
\begin{equation}
\label{hso}
H_{so}=2\Lambda\sum_i\left(S_i^x\cos\phi_i
                          +S_i^z\sin\phi_i\right)\tau_i^y,
\end{equation}
and we use $\lambda=\Lambda/J$ as a free parameter. In order to 
understand the important consequences of the tilting for the interplay 
between spin and orbital degrees of freedom, we consider first the 
idealized structure with $\phi=0$. The coherent spin-and-orbital 
fluctuations lower then the energy due to on-site correlations 
$\langle S_i^x\tau_i^y\rangle<0$. Since these fluctuations do not couple 
to the spin order, no orbital moments can be induced at small $\lambda$
as long as $\phi=0$. 

\begin{figure}
\includegraphics[width=8.7cm]{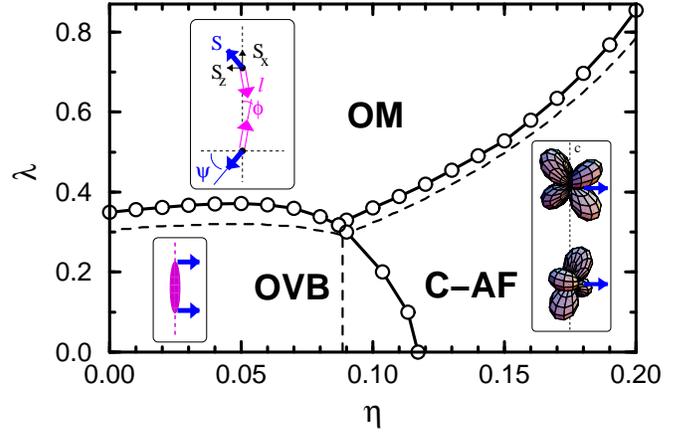}
\vskip -.1cm
\caption{
Phase diagram of the spin-orbital model in the $(\eta,\lambda)$ 
plane at $T=0$, reflecting the competition between 
orbital valence bond (OVB), staggered orbitals (C-AF) and
orbital moment ordering (OM),
as obtained by the ED 
of a four-site embedded chain for $\phi=11^{\circ}$ (circles), and in 
the MFA (dashed lines). Orbital moments in the OM phase (violet arrows) 
induce spin canting (blue arrows) with angle $\psi$. 
(YVO$_3$: $\eta\simeq 0.12$, $\lambda\sim 0.3 - 0.4$ \cite{notept}).
} 
\label{fig:phd}
\end{figure}

Even in the absence of the spin-orbit term ($\lambda=0$), the vanadate 
spin-orbital model (\ref{som}) poses a highly nontrivial quantum 
problem. We obtained first qualitative insight into the possible types 
of magnetic and orbital ordering by investigating the stability of 
different phases within the mean-field approximation (MFA), but 
including the leading orbital fluctuations on FM bonds along the $c$ 
axis. In the absence of Hund's exchange ($\eta=0$), two orbital flavors 
experience an antiferro-orbital (AO) coupling on these bonds due to 
${\hat J}_{ij}^{(c)}$ Eq. (\ref{jc}), but are decoupled within the 
$(a,b)$ planes (${\hat K}_{ij}^{(a,b)}=0$ \cite{noteja}). This 
one-dimensional (1D) system is unstable towards {\it dimerization\/} 
with orbital singlets and FM interactions at every second bond along 
the $c$ axis \cite{She02}, stabilizing the orbital valence bond (OVB) 
phase. The spin interactions ${\hat J}_{ij}^{(a,b)}$ Eq. (\ref{ja}) 
and the intersinglet interactions along $c$ axis are AF. In contrast, 
for large $\eta$ 
more energy is 
gained when the orbital singlets resonate along the $c$ direction, 
giving uniform FM 
interactions in the $C$-AF phase \cite{Kha01}. We determined the quantum 
energy due to orbital fluctuations using the orbital waves found in the 
Schulz approximation, known to be accurate for weakly coupled AF chains 
\cite{Sch96}. Using this approach, the transition from the OVB to $C$-AF 
phase [$\langle S_i^z\rangle=S^z e^{i{\vec R}_i{\vec Q}_C}$ with 
${\vec Q}_C=(\pi,\pi,0)$] takes place at $\eta_c\simeq 0.09$ 
(Fig. \ref{fig:phd}), and the orbital ordering, 
$\langle\tau_i^z\rangle=\tau^z e^{i{\vec R}_i{\vec Q}_G}$ with 
${\vec Q}_G=(\pi,\pi,\pi)$, sets in, promoted by the AO interactions 
${\hat K}_{ij}^{(a,b)}$ Eq. (\ref{ka}).

The ground state changes qualitatively at finite $\lambda$ and $\phi>0$. 
The magnetic moments $\langle S_i^z\rangle$ induce then the {\it orbital 
moments\/}, $2\langle\tau_i^y\rangle=l^z e^{i{\vec R}_i{\vec Q}_A}$, 
which stagger along the $c$ axis with ${\vec Q}_A=(0,0,\pi)$. This 
novel type of ordering with $l_i^z\neq 0$ can be described as a staggered
ordering of complex orbitals $a \pm ib$ (corresponding to $l_i^z=\mp 1$ 
eigenstates) that competes at $\eta>\eta_c$ with the staggered ($a/b$) 
ordering of real orbitals with $\tau^z\neq 0$ in the $C$-AF phase.
Already at small $\lambda$ the orbital moments induce in turn opposite 
to them weak $\langle S_i^x\rangle\neq 0$ moments, lowering the energy 
by $\langle S_i^x\tau_i^y\rangle<0$. In the OM phase favored at large 
$\lambda$ (Fig. \ref{fig:phd}), the spin order has therefore two 
components: $\langle S_i^z\rangle=S^z e^{i{\vec R}_i{\vec Q}_C}$, and 
            $\langle S_i^x\rangle=S^x e^{i{\vec R}_i{\vec Q}_G}$.

An unbiased information about the spin and orbital degrees of freedom 
was obtained by the accurate treatment of quantum effects within the 
exact diagonalization (ED) method. Thereby the coupled spin-orbital 
excitations, terms
$\propto S_i^{\alpha}S_j^{\alpha}\tau_i^{\beta}\tau_j^{\beta}$ in Eq. 
(\ref{som}), are now fully included. We performed ED of four-site chains 
along $c$ axis, both for free and periodic boundary conditions (PBC). 

We were surprised to see that the exact ground state of a 
{\it free chain\/} at $\eta=\lambda=0$ consists indeed in a very good 
approximation of two orbital singlets on the external (12) and (34) 
FM bonds ($\langle\vec\tau_i\cdot \vec\tau_{i+1}\rangle=-0.729$), 
connected by an AF bond (23) with decoupled orbitals 
($\langle\vec\tau_i\cdot \vec\tau_{i+1}\rangle=-0.038$). 
The spin correlations are FM/AF on the external/central bond, 
$\langle\vec S_i\cdot\vec S_{i+1}\rangle=0.95$ and $-1.56$. 
With increasing $\eta$ the AF interaction weakens, the sequence 
of spin multiplets labelled by the total spin $S_t$ is inverted at 
$\eta_c\simeq 0.12$, and the ground state changes from a singlet 
($S_t=0$) to a high-spin ($S_t=4$) state. At finite $\lambda$ no level 
crossing occurs, but the nondegenerate ground state describes a smooth 
crossover in the spin and orbital correlations with increasing $\eta$. 
We verified that several excited states lie within $0.1J$ away from 
the ground state --- all of them would contribute to thermal 
fluctuations already at temperatures $T\simeq 30$ K.

\begin{figure}
\includegraphics[width=6.8cm]{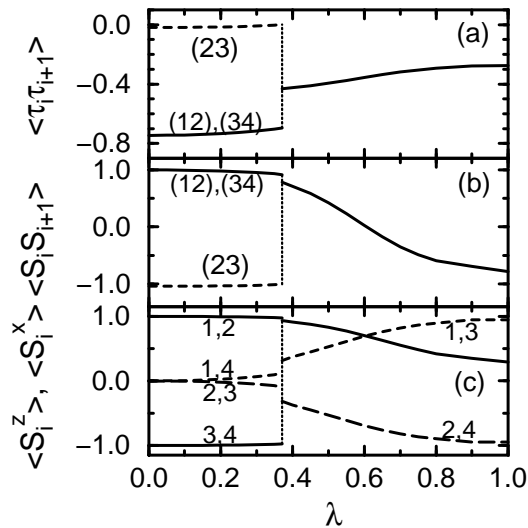}
\vskip -.1cm
\caption{
Pair correlations along $c$ axis:
(a) orbital $\langle{\vec\tau}_i\cdot{\vec\tau}_{i+1}\rangle$, 
(b) spin $\langle {\vec S}_i\cdot{\vec S}_{i+1}\rangle$, and 
(c) spin components $\langle S_i^z\rangle$ (full lines) and 
                    $\langle S_i^x\rangle$ (dashed, long-dashed lines), 
as functions of $\lambda$ for the OVB ($\lambda<\lambda_c\simeq 0.37$) 
and the OM ($\lambda>\lambda_c$) phase, found by the ED at $\eta=0.07$. 
FM/AF bonds $(ij)$ in the OVB phase are shown by solid/dashed 
lines in (a) and (b).}
\label{fig:cor}
\end{figure}

We simulate a {\it cubic system\/} by including infinitesimal 
symmetry-breaking dimerization field which favors the orbital singlets 
on the bonds (12) and (34) in a cluster with the PBC, embedded within 
one of three phases stable in the MFA (Fig. \ref{fig:phd}), with 
mean-fields determined self-consistently in each phase. All phases are 
characterized by finite magnetic moments $\langle S_i^z\rangle$, either 
staggered pairwise (OVB phase), or aligned ($C$-AF phase) along $c$ 
axis, and weak $\langle S_i^x\rangle$ moments. In addition, the orbital 
ordering ($l^z\neq 0$) appears in the OM phase, while the orbitals 
stagger ($\tau^z\neq 0$) in the $C$-AF phase. 

\begin{figure}
\includegraphics[width=6.8cm]{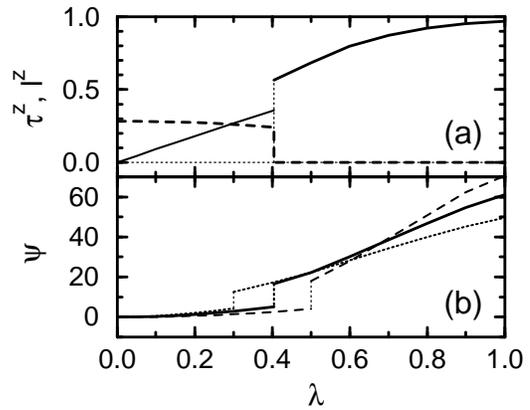}
\vskip -.1cm
\caption{
Crossover from the $C$-AF to the OM phase for increasing $\lambda$
at $\eta=0.12$:
(a) orbital order parameters: $\tau^z$ (dashed line), 
    and $l^z$ (solid lines) at $\phi=11^{\circ}$;
(b) spin canting angle $\psi=\arctan(S^x/S^z)$ (see Fig. 
    \protect\ref{fig:phd}), as obtained for the tilting: 
    $\phi=5^{\circ}$, $11^{\circ}$ and $20^{\circ}$ 
    (dashed, solid and dotted line). }
\label{fig:mom}
\end{figure}

By computing the energies of different phases we obtained the phase
diagram that confirms the qualitative picture extracted from the MFA
(Fig. \ref{fig:phd}). All transitions are accompanied by reorientation 
of spins (Figs. \ref{fig:cor} and \ref{fig:mom}). The orbital and spin 
fluctuations change only weakly at small values of $\lambda$, weak 
$\langle S_i^x\rangle$ moments are ordered pairwise on the bonds (12) 
and (34), and the spin correlations on the intersinglet bonds are 
almost classical (Fig. \ref{fig:cor}). These features show that the 
spins and orbitals are almost decoupled and the OVB phase is robust. 
At $\lambda>\lambda_c$ the on-site correlations 
$\langle S_i^x\tau_i^y\rangle<0$ dominate, while the orbital 
fluctuations are suppressed, and the correlation functions 
$\langle {\vec\tau}_i\cdot {\vec\tau}_{i+1}\rangle$ approach the 
classical value $-\frac{1}{4}$. As the staggered spin moments 
$\langle S_i^x\rangle$ are induced, the spin correlations
$\langle {\vec S}_i\cdot{\vec S}_{i+1}\rangle$ become soon AF within 
the OM phase. In this regime the spins follow the spin-orbit coupling 
$\lambda$, and the FM interaction $J_c$ is frustrated.  

At the transition from the $C$-AF to the OM phase the orbital ordering 
changes from staggered {\it real orbitals\/} ($\tau^z\neq 0$) to 
staggered {\it orbital moments\/} ($l^z\neq 0$), as shown in Fig. 
\ref{fig:mom}(a). As a precursor effect of the forthcoming OM phase, 
the orbital moments $l_i^z$ are induced already in the $C$-AF phase by 
increasing $\lambda$. The transition to the OM phase is accelerated by 
the increasing tilting angle $\phi$ [Fig. \ref{fig:mom}(b)]. Also the 
spin correlations change here discontinuously at the transition (not 
shown), similar to the OVB/OM transition [Fig. \ref{fig:cor}(b)]. The 
staggered spin components in the OM phase $\langle S_i^x\rangle$ are 
similarly large to those shown in Fig. \ref{fig:cor}(c) for smaller 
$\eta$, and the spin canting angle $\psi$ approaches 
$\frac{\pi}{2}-\phi$ in the regime $\lambda\gg 1$. For realistic 
parameters for YVO$_3$: $J\sim 30$ meV \cite{Kha01}, 
$\eta\sim 0.12$ (estimated with $J_H=0.64$ eV and intraorbital 
element $U=5.5$ eV for V$^{2+}$ ions \cite{Zaa90}), 
and $\lambda\sim 0.3 - 0.4$ (considering $\Lambda\simeq 13$ meV for 
free V$^{3+}$ ions \cite{Abrag}), one finds a competition between the 
staggered $a/b$ orbital order parameter ($\tau^z\sim 0.25$) 
and the orbital magnetic moments ($l^z\sim 0.30-0.35$). 
This reflects the interplay between intersite SE and on-site spin-orbit
couplings, and hence orbital and spin moments are not
collinear (except for large $\lambda$ values), --- in contrast to the
conventional picture where orbital moments induced by $\lambda$ coupling
are antiparallel to spin, as suggested e.g. for V$_2$O$_3$ \cite{Tan02}.

\begin{figure}
\includegraphics[width=7.1cm]{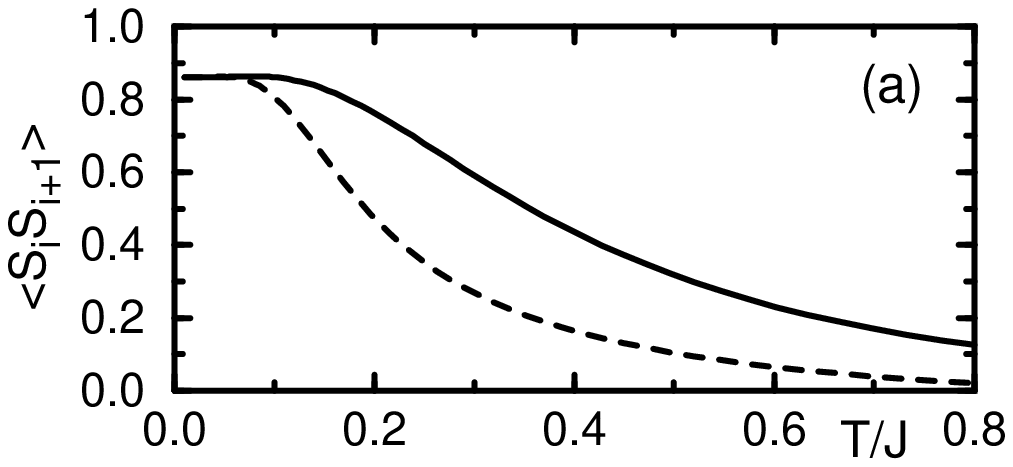}
\includegraphics[width=6.8cm]{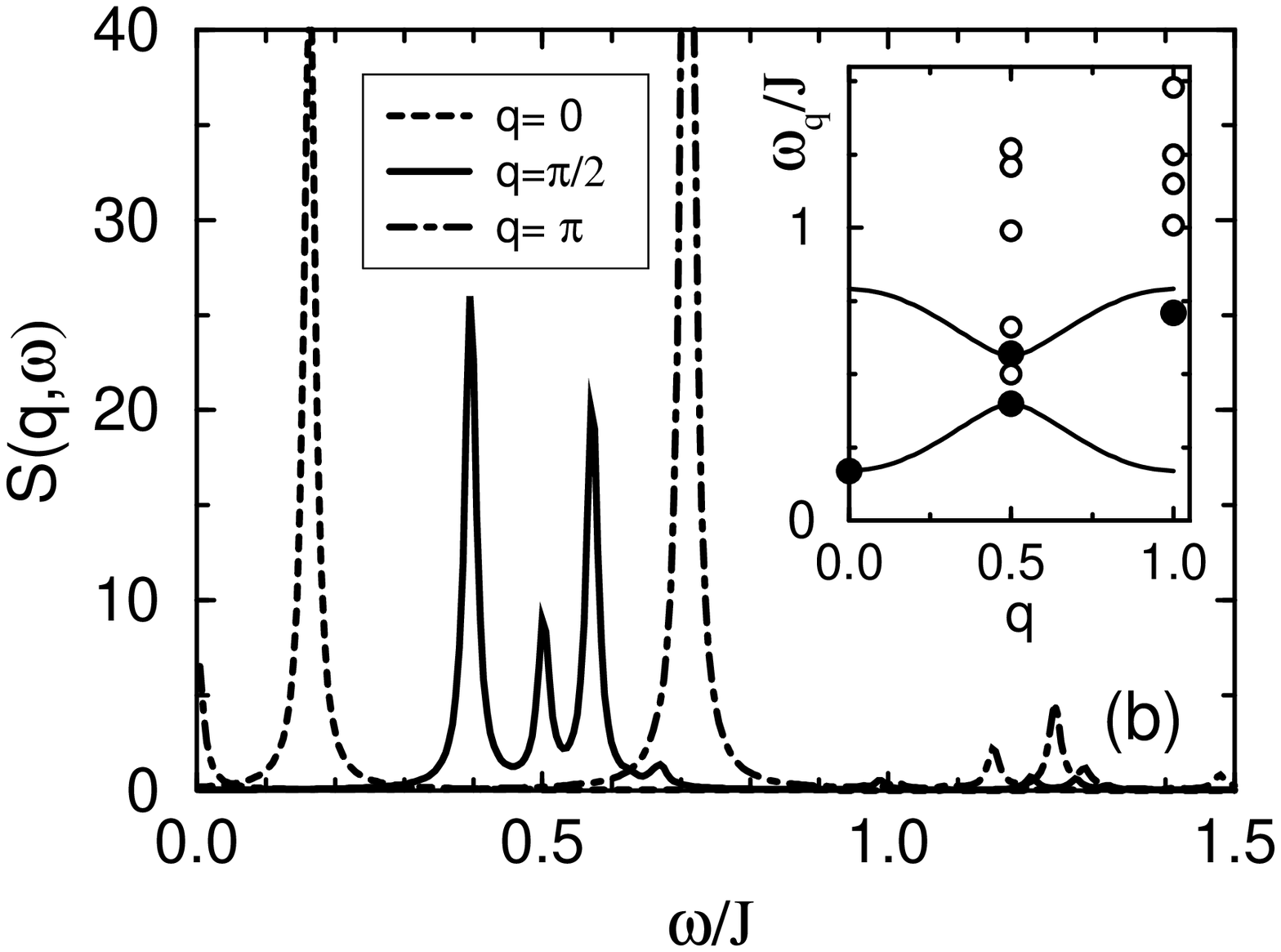}
\vskip -.1cm
\caption{
Dimerization in the $C$-AF phase at $T>0$:
(a) spin-spin correlations $\langle{\vec S}_i\cdot{\vec S}_{i+1}\rangle$
    on strong and weak FM bonds (solid and dashed line);
(b) spin response $S({\bf q},\omega)$ in the dimerized $C$-AF phase for 
    $q=0, \pi/2, \pi$. Inset: filled (open) circles indicate
    strong (weak) features in $S({\bf q},\omega)$; lines show the 
    fitted spin-wave dispersion. 
Parameters: $\eta=0.12$, $\lambda=0.4$. 
}
\label{fig:dim}
\end{figure}

Finally, we turn to finite temperatures. While the $C$-phase cannot 
dimerize at $T=0$, it dimerizes at finite $T$ due to the intrinsic 
instability towards alternating orbital singlets \cite{Sir02}.
The spin correlations $\langle {\vec S}_i\cdot{\vec S}_{i+1}\rangle$ 
[Fig. \ref{fig:dim}(a)], found using open boundary conditions, 
alternate between strong and weak FM bonds due to the orbital Peierls 
dimerization, $2\delta_{\tau}=
|\langle {\vec\tau}_i\cdot {\vec\tau}_{i+1}\rangle
-\langle {\vec\tau}_i\cdot {\vec\tau}_{i-1}\rangle|$, which has a 
distinct maximum at $T\simeq 0.24 J$ for $\eta=0.12$. Consistent with 
our discussion above, the modulation of
$\langle{\vec S}_i\cdot{\vec S}_{i+1}\rangle$ vanishes 
in the C-phase at low $T$.
  
Up to now, the sole experimental evidence for dimerization of the 
$C$-phase is the splitting of FM spin waves in the neutron scattering 
study of Ulrich {\it et al.} \cite{Ulr02}. Fig. 4(b) shows the 
dynamical spin structure factor $S(q,\omega)$ \cite{Aue94} obtained by 
exact diagonalization of a 4-site cluster with PBC at $T=0$, assuming 
the same orbital dimerization  
$\langle{\vec\tau}_i\cdot{\vec\tau}_{i+1}\rangle$ as found above for 
$T/J=0.25$. At $q=\pi/2$ we observe a splitting of the spin wave similar 
to experiment. The finite energy of the $q=0$ mode results from the 
$\lambda$-term and the mean-field coupling to neighbor chains.
Additional features seen in $S(q,\omega)$, e.g. for $q=\pi$ at
$\omega \sim 1.25 J$, we attribute to the coupling to orbital
excitations. The spin-wave energies can be fitted by a simple 
Heisenberg model with two FM coupling constants: $J_{c1}=5.7$,
$J_{c2}=3.3$ meV, and a small anisotropy term, as shown in the inset 
(solid lines). Although these values are strongly reduced 
by spin-orbit coupling 
Eq. (\ref{hso}), they are still larger than those extracted from the 
spin waves in YVO$_3$: $J_{c1}^{exp}=4.0$ and $J_{c2}^{exp}=2.2$ meV 
at $T=85$ K \cite{Ulr02}. We attribute this overestimate of exchange 
interactions to quantum fluctuations involving the occupancy of 
$xy$-orbitals; this will be treated elsewhere.

Summarizing, we have shown that the spin-orbit coupling $\Lambda$ 
competes with Hund's exchange in the spin-orbital model for cubic 
vanadates. It leads to a new orbital moment ordered phase at large 
$\Lambda$ and can explain qualitatively the spin canting and large 
reduction of magnetization in the $C$-phase  at smaller $\Lambda$. 
We argue that the 1D orbital Peierls instability observed recently 
in the $C$-AF phase of YVO$_3$ \cite{Ulr02} (along the $c$ axis) 
emerges from a combination of quantum effects due to orbital moments 
at $\Lambda>0$, and thermal fluctuations which favor dimerized orbital 
and spin correlations. 

We thank B. Keimer and C. Ulrich for stimulating discussions. 
This work was supported by the Committee of Scientific Research (KBN), 
Project No.~5 P03B 055 20.



\begin{thebibliography}{}
\bibitem{noteja} Spins are AF, so ${\hat J}_{ij}^{(a,b)}$ do not 
                 contribute either.
\bibitem{notept} Below $T_{N1}=$77 K, the $C$-AF phase of YVO$_3$ 
                 changes to the $G$-type AF spin structure 
		 \protect\cite{Ren00}. As we have argued 
		 \protect\cite{Kha01}, the $G$-AF phase is stabilized 
		 at low temperature by the JT effect, and the phase
                 transition at $T_{N1}$ is promoted by the larger 
		 (spin and orbital) entropy of the $C$-AF phase.
\end{thebibliography}
\end{document}